\documentclass{article}
\usepackage{todonotes}
\usepackage{natbib}
\usepackage{wrapfig}
\usepackage{hyperref}
\usepackage{geometry}
\usepackage{authblk}

\usepackage{ccicons}          
\title{Circumvention by design - dark patterns in cookie consents for online news outlets}

\author{Than Htut Soe}
\author{Oda Elise Nordberg}
\author{Frode Guribye}
\author{Marija Slavkovik}
\affil{Department of Information Science and Media Studies, University of Bergen, Norway}
\affil{\{than.soe, oda.nordberg, frode.guribye, marija.slavkovik\}@uib.no}
\date{}                     
\begin{document}

\maketitle

\begin{abstract}
To ensure that users of online services understand what data are collected and how they are used in algorithmic decision-making, the European Union's General Data Protection Regulation (GDPR) specifies informed consent as a minimal requirement. For online news outlets consent is commonly elicited through interface design elements in the form of a pop-up. We have manually analyzed 300 data collection consent notices from news outlets that are built to ensure compliance with GDPR. The analysis uncovered a variety of strategies or dark patterns that circumvent the intent of GDPR by design. We further study the presence and variety of these dark patterns in these ``cookie consents'' and use our observations to specify the concept of dark pattern in the context of consent elicitation. 
\end{abstract}



 \maketitle

\section{Introduction}

Following the adoption of the General Data Protection Regulation  (GDPR), more than 60 percent of popular websites in Europe use pop-ups to elicit informed consent to their privacy policy and their use of cookies to collect and anlyze data about their audience \citep{UtzDFSH2019}. While this ostensibly annoying feature has been the most notable implication of GDPR for the regular user of the web, the last year it has also received critical attention from the research community  \citep{UtzDFSH2019,MatteBS2019,NouwensLVKK20}.
In practice, the regulation of data collection and privacy for online services requires, not only that the process of automation and augmentation \citep{Garfinkel2009} is adequately managed, but also that the services offer a carefully designed interface. These interfaces are not a neutral conduit but can help or hinder the user in acting in their own best interest \citep{Leitao:2018}.

The term dark patterns \citep{Gray:2018}, has been coined\footnote{The neologism, dark pattern, was coined by user experience designer Harry Brignull in 2010.} to identify ``instances where designers use their knowledge of human behavior (e.g., psychology) and the desires of end users to implement deceptive functionality that is not in the user's best interest" \citep{Gray:2018}. Dark patterns can be used to circumvent the intent of GDPR, by design \citep{ForbrukerradeNO}; even when that service is explicitly required to protect the interest of the user, as is the case with the elicitation of informed consent \citep{DucatoE2019}. We study the extent of use of dark patterns specifically in the case of the designs used to elicit informed consent for online news services. We use our findings to argue, in line with \citet{NouwensLVKK20}, that {\bf any regulation of a computational system that aims to protect user rights should be accompanied by a regulation of user interface design.}

Today regulation is put in place to ease the information asymmetry by giving control to individuals over their personal data: The European Union's General Data Protection Regulation (GDPR) that  has been in force since May 25, 2018 \citep{GDPR};
and The California Consumer Privacy Act (CCPA) of 2018 that has been in force since January 1, 2020 \citep{CCPA}. 
However, this regulation includes very little guidelines, if at all, regarding the user interface that communicates the rights to users and operationalises their right to control. 

In response to the GDPR and CCPA regulations, websites display so called {\em cookie consent notices}.  We conducted a survey of online news outlets and manually collected and analyzed the cookie consent notices deployed. We were specifically interested in identifying the prevalence in use of dark patterns as described by \citet{Gray:2018}. We also collected information on the complexity of the consent notice design in general. We cannot attest to the intention behind the  design choices that we encountered, however since dark patterns do ``nudge'' a user towards a particular behaviour \citep{Gray:2018},  we use the concept of  dark  patterns as a proxy to bad, if not malevolent,  design that leads to an unethical UI. 
 
Specifically, we manually analyzed 300 consent notifications from Scandinavian and English language news outlets and found that all employ some level of unethical practices. We  chose  online news outlets for two reasons. News readers span across demographics\footnote{88\% of read news are  accessed online according to 75.000 people surveyed across 38 countries \citep{Reuters2019}.} and cookie consent notices here can be expected to be designed to be usable by all demographic (as opposed to, for example, specialists in a given area). Furthermore, news outlets do not primarily trade in user data, like for example social media services, and one of their main roles is to be a service of society.  As a societal service, we expect that there is an incentive to avoid consent designs that manipulate their readers towards giving consent to avoid the impression that the outlet is manipulating public opinions rather than reporting truthfully the news.  On the other hand  most news outlets rely on online advertising in their business model, so the incentive to collect user data is not entirely removed. We initially focused on news outlets in Scandinavia, and then included in our collection  international news outlets with content  in English. Our motivation to focus on Scandinavia was both practical and strategic. The authors are able to understand the Scandinavian languages and have access to an exhaustive list of Scandinavian news outlets. The Scandinavian media enjoys a reputation of trust and freedom of speech \citep{Rank2020} and we could expect to see the same values respected in complying with GDPR. 

In the past year there had been at least three works that explicitly study how consent elicitation is implemented \citep{UtzDFSH2019,MatteBS2019,NouwensLVKK20} with \citet{NouwensLVKK20} specifically looking into dark patterns. These researchers conducted their analysis by crawling a selected choice of websites: e-commerce \citep{UtzDFSH2019}, UK \citep{NouwensLVKK20} and European \citep{MatteBS2019}. In contrast we focus on design aspects of dark patterns in consent notices and  overall ease of denying consent expressed. In contrast to \citep{UtzDFSH2019,MatteBS2019,NouwensLVKK20} our data set is smaller, however we study features that are very difficult to detect automatically. We outline our contribution as follows. 

\paragraph{Contribution.}   
\begin{itemize}
    \item A rich manually collected data set of 300 cookie consent notices. 
    \item An analysis of features of consent notices that cannot be easily detected automatically. 
    \item A set of eight new dark patterns that refine the dark pattern types currently identified for the consent notices design context.
\end{itemize}

This paper is structured as follows. In Section~\ref{sec:related} we introduce relevant preliminary information and position our paper within the related work. In Section~\ref{sec:collected} we describe our data collection process. In Section~\ref{sec:found} we report the dark patterns observed both the existence and the subsequent study on identifying specific dark pattern types. In Section~\ref{sec:complex} we describe further observations of consent notice design features that contribute to how easy or hard it is to deny consent for data collection.  In Section~\ref{sec:refine} we pull on our analysis and the related work to propose how the regulation of consent notice design can be improved by proposing eight new dark pattern types specific for this domain of user interfaces. Lastly, in Section~\ref{sec:summary} we summarise our findings, outline the limitations of our work and discuss future work.

\section{Background and related work}\label{sec:related}

We are here concerned specifically with the design of consent elicitation website elements, which are sometimes called cookie consent notices. These elements are implemented as a pop-up or as a banner or panel that is part of the website. Cookie consent notices are typically implement to demonstrate compliance with the GDPR and CCPA regulations. We do not intend to analyse in detail the legal aspects of consent elicitation that these regulations prescribe, but introducing some detail is necessary.

  GDPR refers \citep{GDPR} to interaction between humans and computational systems in which human consent for the operation of the system is elicited; specifically article (32):
  \begin{description}	
 \item[(32)]	Consent should be given by a clear affirmative act establishing a freely given, specific, informed and unambiguous indication of the data subject's agreement to the processing of personal data relating to him or her, such as by a written statement, including by electronic means, or an oral statement. This could include ticking a box when visiting an internet website, choosing technical settings for information society services or another statement or conduct which clearly indicates in this context the data subject's acceptance of the proposed processing of his or her personal data. Silence, pre-ticked boxes or inactivity should not therefore constitute consent. Consent should cover all processing activities carried out for the same purpose or purposes. When the processing has multiple purposes, consent should be given for all of them. If the data subject's consent is to be given following a request by electronic means, the request must be clear, concise and not unnecessarily disruptive to the use of the service for which it is provided.
 \end{description}

 The CCPA enlists the rights of a user, such as for example ``1798.100. (a) A consumer shall have the right to request that a business that collects a consumer's personal information disclose to that consumer the categories and specific pieces of personal information the business has collected.'' However, the CCPA does not mandate how the user is to exercise such rights.

We are explicitly looking at how easy or hard the design of cookie consent notices makes the elicitation for a user. We consider two types of proxies for ``ease of consent": the use of dark patterns and the interaction complexity of the consent notice. In the later category we look at choice hierarchies, how many clicks it takes a user to withhold consent compared to the effort involved to grant it, and whether there is uniformity of language used, e.g., to refer to different consent options  and cookie categories. 

Dark patterns as a term was introduced by  Harry Brignull who defines them as ``tricks used in websites and apps that make you do things that you didn't mean to" \citep{Brignull2010}. The concept of Dark Patterns in user experience design  was  refined by \citet{Gray:2018} who specified five different types of dark patterns: nagging, obstruction, sneaking, interface interference, and forced action.  We include the description of each of these patterns as given  by \citet{Gray:2018} in Table~\ref{tab:dark}.
 \begin{table}[t!]
    \centering
    \resizebox{0.9\textwidth}{!}{
    \begin{tabular}{p{1.8cm}|p{10cm}}\\
    Name  & Description\\\hline
   Nagging     & A minor redirection of expected functionality that may persist over one or more interactions. Nagging often manifests as a repeated intrusion during normal interaction, where the user's desired task is interrupted one or more times by other tasks not directly related to the one the user is focusing on. \\\hline
     Obstruction    & Impending a task flow, making an interaction more difficult than it inherently needs to be with the intent to dissuade an action. Obstruction often manifests as a major barrier to a particular task the user may want to accomplish.\\\hline
     Sneaking & An attempt to hide, disguise, or delay the divulging of information that has relevance to the user. Sneaking often occurs in order to make the user perform an action they may object to if the had the knowledge.\\\hline
     Interface $\;\;$ interference & Any manipulation of the user interface that privileges specific actions over others, thereby confusing the user or limiting discoverability of important action possibilities. Interface interference manifests as numerous individual visual and interactive deceptions.\\\hline
     Forced \linebreak action & Any situation in which users are required to perform a specific action to access (or continue to access) specific functionality. This action may manifest as a required step to complete a process, or may appear disguised as an option that the user will greatly benefit from.
    \end{tabular}}
    \caption{Dark pattern types of \citep{Gray:2018} and their definitions}
    \label{tab:dark}
\end{table}

 It was expected that a regulation such as GDPR would lead to dark pattern proliferation \citep{Paternoster2018}. It is therefore unsurprising that in the past year there are works like ours that looked specifically in how consent elicitation of cookie consent notices was executed. 

\citet{UtzDFSH2019} have conducted a field study of consent notices on a live (e-commerce) website to identify how does the design of the notice influence the user decision to accept the website cookies. They studied 80,000 unique users. Specifically they looked into the relative position of the notice, use of nudging and the presence of a privacy link (that explains in detail how data is collected and used) and showed that small UI design decisions substantially impact whether and how people interact with  cookie consent notices. One of their experiments indicated that nudging via interface interference (highlighting Accept button in a binary choice with decline) and pre-selected choices for different uses of cookies has a strong impact of whether the users accept the third-party cookies.

\citet{NouwensLVKK20} also performed a study on the impact of various designs of consent notices, user interface design nudges and level of granularity of options. They focus on Consent Management Platforms (CMP), which are a service offered by third parties to  website owners to help them outsource regulatory compliance and as part of their service promise compliance with regulation. \citet{NouwensLVKK20}  scraped the designs of the five most popular CMPs on the top 10,000 websites in the UK which has yielded 680 notices. Specifically they were looking for the presence of three features: whether consent is explicit, if ease of acceptance is the same as rejection  (by checking whether accept is the same widget (on the same hierarchy) as reject),and whether pre-ticked boxes are present. They found that only 80 out of the 680 notices satisify all three conditions.

\citet{NouwensLVKK20} also ran a user analyses, on 40 participants, on the effect the consent notice design has on whether consent is given. They found that there was an approximate 22\% of increase in acceptance when the opt-out option was ``hiden'' behind the initial notice (at least two clicks are needed to opt out).  
  
\citet{MatteBS2019}  also focus on CMPs, they looked at those who comply with  Europe's Transparency and Consent Framework (TCF). They ran two automatic and
semi-automatic crawl campaigns to detect suspected regulatory violations. Specifically they were looking at  consent stored behind the user interface of  cookie notices. They crawled 28257 European websites (in 2019) and detected  suspected violations in  1426 of the visited websites. \citet{MatteBS2019}  specifically looked at whether consent was stored before the user made the choice, whether the notice offers a way to opt out, whether there were pre-selected choices and lastly if the choice that the user had made was respected at all. They found that ``141 websites register positive consent even if the user has not made their choice; 236 websites nudge the users towards accepting consent by pre-selecting options; and 27 websites store a positive consent even if the user has explicitly opted out." \citep{MatteBS2019}. 
 
 \citet{NouwensLVKK20} have constructed a browser extension, Consent-O-Matic \citep{ConMAT}, which allows users to set their consent preferences once and have them automatically applied to visited websites. \citet{MatteBS2019}  also built and published a browser extension called Cookie Glasses \citep{CookieGlasses} to enable users to see if consent stored by CMPs corresponds to their choice. Both extensions are open source and freely downloadable.

It should be noted that the ambiguity of the term ``informed consent"  is in itself an issue \citep{NIK2018}. \citet{NouwensLVKK20} make considerable effort to identify the features of the legal understanding of consent in the laws of the European Union. They work with the definition of  art 4(11) of the GDPR: "any freely given, specific, informed and unambiguous indication of the data subject's wishes by which he or she, by a statement or by a clear affirmative action, signifies agreement to the processing of personal data relating to him or her".  \citet{UtzDFSH2019}  conclude that opt-out consent banners are unlikely to produce intentional or meaningful expression of consent \citep{UtzDFSH2019}.

 \section{Collected data on Cookie  notice implementations}\label{sec:collected}
  The study was done in two passes. First, consent notices from 300 news outlets with content in Scandinavian languages or English were manually collected and analyzed in the period of July 1-30, 2019.   The full list of accessed link and collected data can be accessed at \url{https://github.com/anoauthor101/anorepo101}. In April (3 months after CCPA became effective) we re-visited the original 300 web-pages and attempted to identify specifically which dark patterns are present in the consent notices. We first describe the first data collection. 
 
Cookie consent notices from 250  online news outlets  and 50 online magazines were collected. Each link was accessed from a web browser with private browsing mode enabled on a laptop. Due to language restriction of the collectors, the  language of the page had to be in English or a Scandinavian language. Only websites that displayed a consent notice were included in the survey. Many Scandinavian news outlets used the same design (third party application) for the consent notice, e.g. Schibsted owns several news outlets and use the same design;  only one example was retained from these.    We aimed for this  number of examples to make manual overview and data analysis feasible, while at the same time have a large enough number of examples to draw meaningful conclusions.  In total, sources from 42 different countries were considered out of which: 33 from Denmark, 34 from Norway, 33 from Sweden, 68 from the UK and 54 from the USA.  In addition to manually identifying the features of cookie consent notices we also collected raw data. The raw data was collected to make our observations verifiable, as websites frequently change, and for possible future research use. The raw data for each surveyed link contains: 
\begin{itemize}
\item In an image format, a screenshot of the first GDPR/data consent notice. If the initial consent notice opened another pop-up window, a screenshot of that was collected as well. If there was a redirect to another page, the page was saved as a pdf file and the text was extracted and saved in a text file. sends the user to other web-pages, screenshots of those were collected as well. 
\item The text of the consent statement. Namely, the text informing the user what they consent to (e.g. information about settings/options, privacy policy, or similar). This was collected as either a PDF (usually privacy policy or similar) or as a text file.  
\item For the consent notices that offer  specific cookie information (companies, cookie names and purposes), this information was also saved as a text file.  
\item An image of the nudging post-consent-denial reconsideration request, if such appears. 
\end{itemize}

Due to space restrictions, we present here the highlights of our analysis and invite the reader to further inspect our data  at \url{https://github.com/anoauthor101/anorepo101}. For each consent notice we identified the following: 

\begin{itemize}
\item{\bf Existence and variety of dark patterns.} Using the dark pattern classification of \citep{Gray:2018}, in July 2019 we observed that all surveyed outlets can be considered to exhibit dark pattern use. This number has slightly changed when we revisited the websites in 2020 and in April 2020 we observed that 3 of our web-pages have no dark patterns, two are geoblocking European visitors and   while 16 have removed the consent notice.  
\item {\bf Possibility for the user to not give consent.} We analyzed whether the consent notice includes an option to not give consent. If an opt-out is present, what text identifies it? Is the widget for the consent option of the same type and complexity as the widget for the  `no consent' option? We checked if the website information is accessible when consent is withheld and  we also checked if the consent notice  explicitly used nudging by asking the user to reconsider after they have chosen to not give consent. 
 \item{\bf Location of the consent notice on the screen.}  
 We observed the vertical position, horizontal alignment and the height of the consent box. The vertical location is identified as either top, bottom or middle of the screen. Horizontal alignment are entire (filling the whole width of the page), left corner or right corner. The height of the box is classified into less than one third, around one third and half of the browser screen height. We also explored whether the content is accessible while the notice is still active or consent is denied.
\item{\bf Complexity of the consent notice.} As a proxy for complexity we used the number of clicks necessary to withhold consent, and the number of words used to explain data uses.  We also considered if the notice describes or links to a website that explains what cookies are, how cookies are used and list their  cookies.
\end{itemize}

\section{Findings}\label{sec:found}
We here summarise our observations of the 300 analysed webpages. We first give examples of the dark patterns described in  \citet{Gray:2018} that our data collectors encountered in July 2019 and then discuss the specific types and frequency of occurrence  that our analysis in April 2020 revealed. Of the five pattern types defined by \citet{Gray:2018}, we observed that
   obstruction and interface interference were the dominant ones in our collected consent notices. We consider as forced action the instances when  a website does not include a consent elicitation but just informs the user that their data is collected. 

 \begin{wrapfigure}{l}{0.4\textwidth}{
\centering
\includegraphics[width=0.4\textwidth]{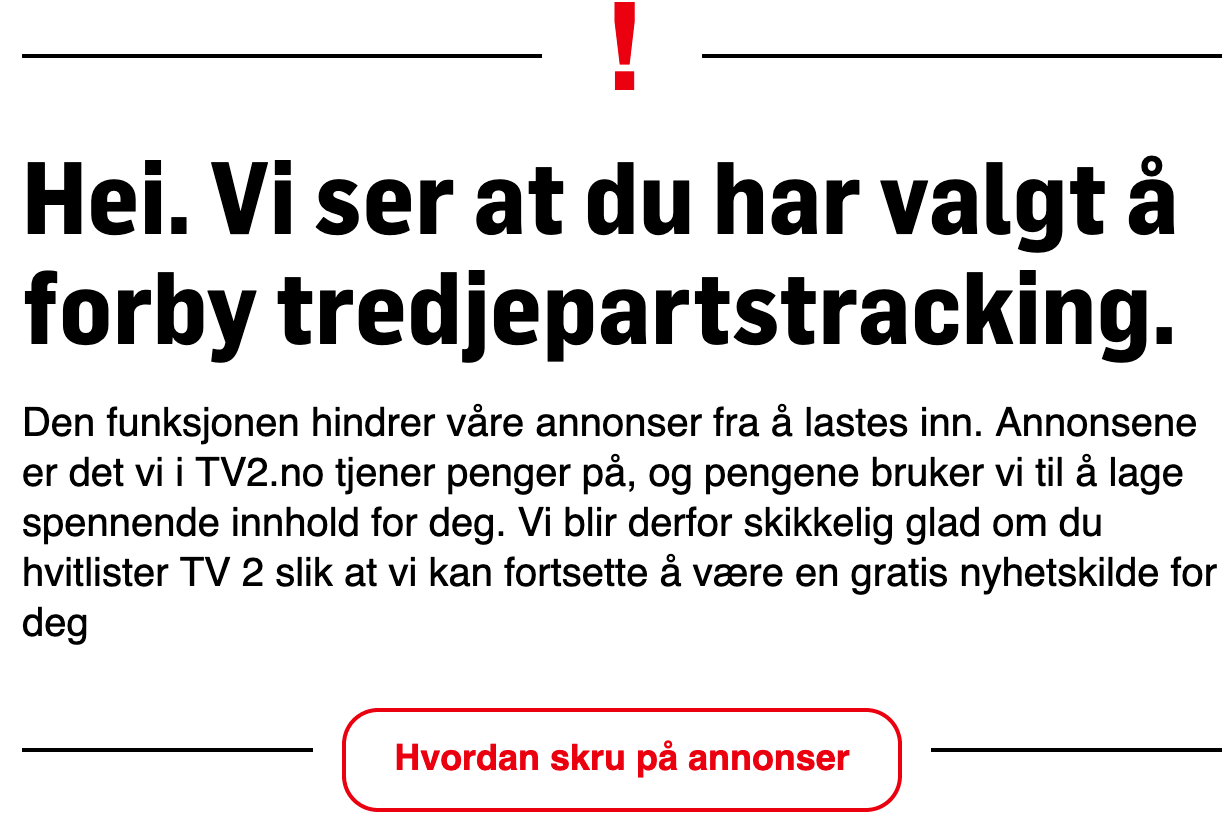}
\caption{\footnotesize An example of nagging.} \label{fig:nagging}
 }
\end{wrapfigure}
\subsection{Existence and examples}
\paragraph{Nagging.}
We observed cases where after consent was denied, the website tried once again to push the visitor to change their mind by displaying a ``Are you sure" notice. Figure~\ref{fig:nagging} gives an example of such a nagging notice displayed from TV2.no after the user opts out from all tracking cookies. The notice reads: Hi. We see that you chose to opt out of third party tracking. This function will prevent our advertisements from downloading. Advertisements are how we in TV2.no earn money and the money we use to make exciting content for you. We would therefore be really happy if you whitelist TV2 so that we can continue to be your free news source.

\begin{figure}[h!]
\centering
\includegraphics[width=0.9\textwidth]{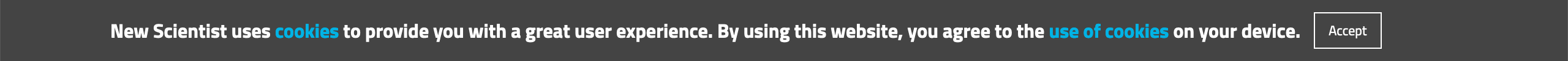}
\caption{\footnotesize An example of obstruction.  } \label{fig:obstruction}
\end{figure}

\paragraph{Obstruction}
Obstruction is both the most prevalent dark pattern and the one most difficult to characterize precisely since a large portion of surveyed websites make the  option of controlling the cookie use and opting out of consent difficult. The most common obstruction pattern is hiding the option to deny consent in a page separate from the consent notice and behind obfuscating text such as ``find out more". Figure~\ref{fig:obstruction} gives an example of such an obstruction encountered at \url{NewScientist.com}. The user can opt out of consent only by making browser adjustments which they can find about after following the ``use of cookies" link. 

\begin{figure}[h!]
\centering
\includegraphics[width=0.9\textwidth]{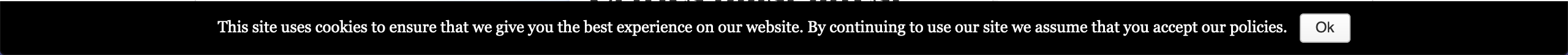}
\includegraphics[width=0.9\textwidth]{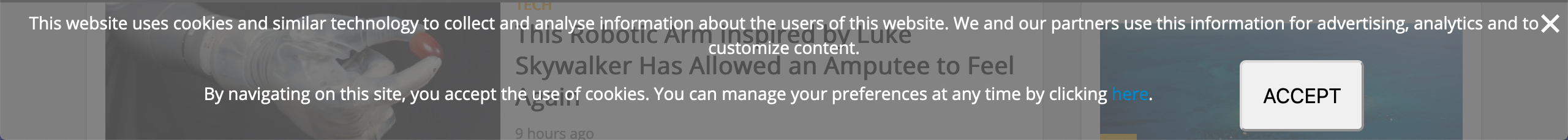}
\caption{\footnotesize Two examples of sneaking.} \label{fig:sneakng}
 
\end{figure}

\paragraph{Sneaking} Sneaking is difficult to separate from obstruction. Several implementations of cookie consent do not give the impression that the user has option to give or deny consent, as was the example with the consent given in Figure~\ref{fig:obstruction}.  This dark pattern can be most clearly seen in widgets that state ``by continuing to use our site, we assume you accept our policies''. Two examples are given in Figure~\ref{fig:sneakng}.  Top figure: the consent notice from NewsInEnglish.no can be seen to be in  violation of GDPR regulation specifically by skirting `` Silence, pre-ticked boxes or inactivity should not therefore constitute consent."  Bottom figure: the consent notice from \url{ScienceAlert.com}.
 
\begin{figure}[h!]
\centering
\includegraphics[width=0.4\textwidth]{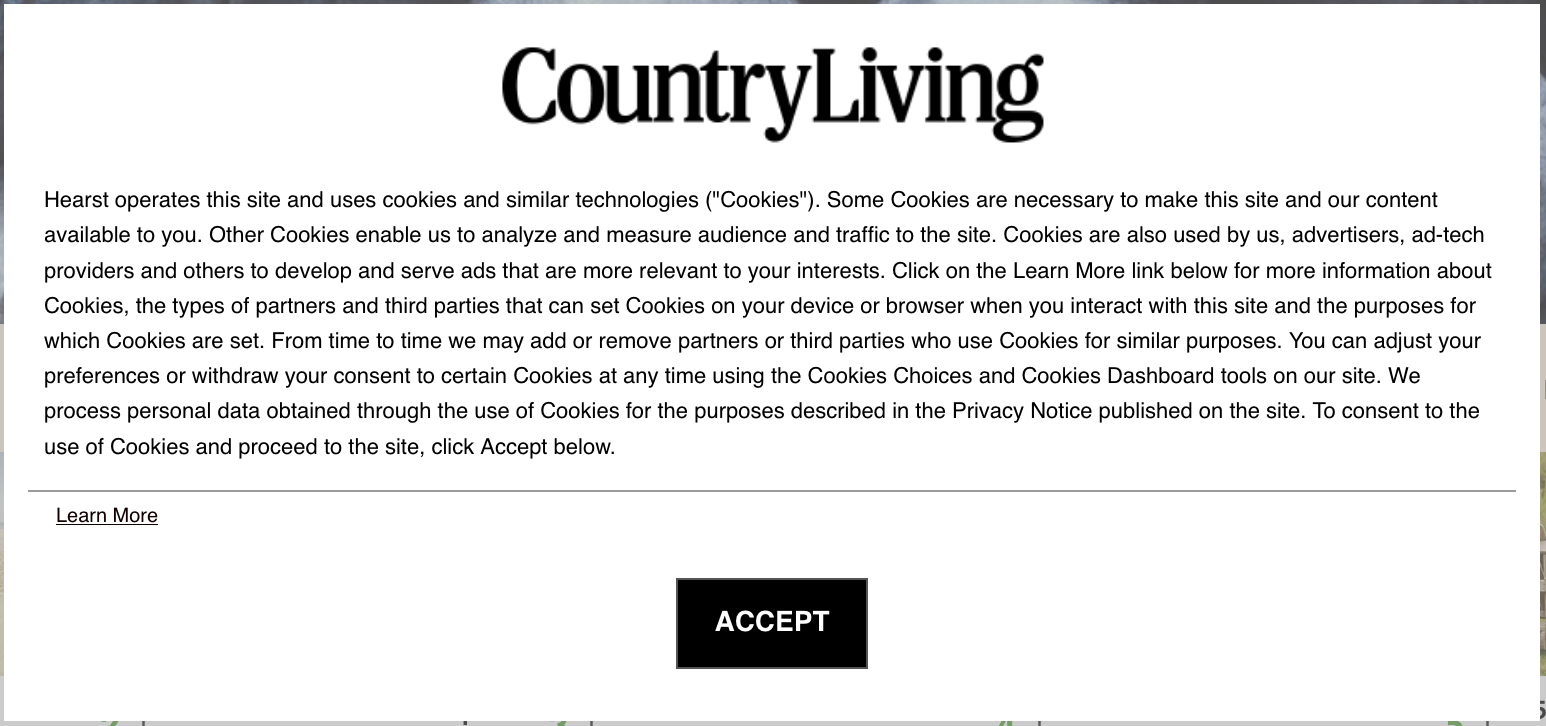}
\includegraphics[width=0.4\textwidth]{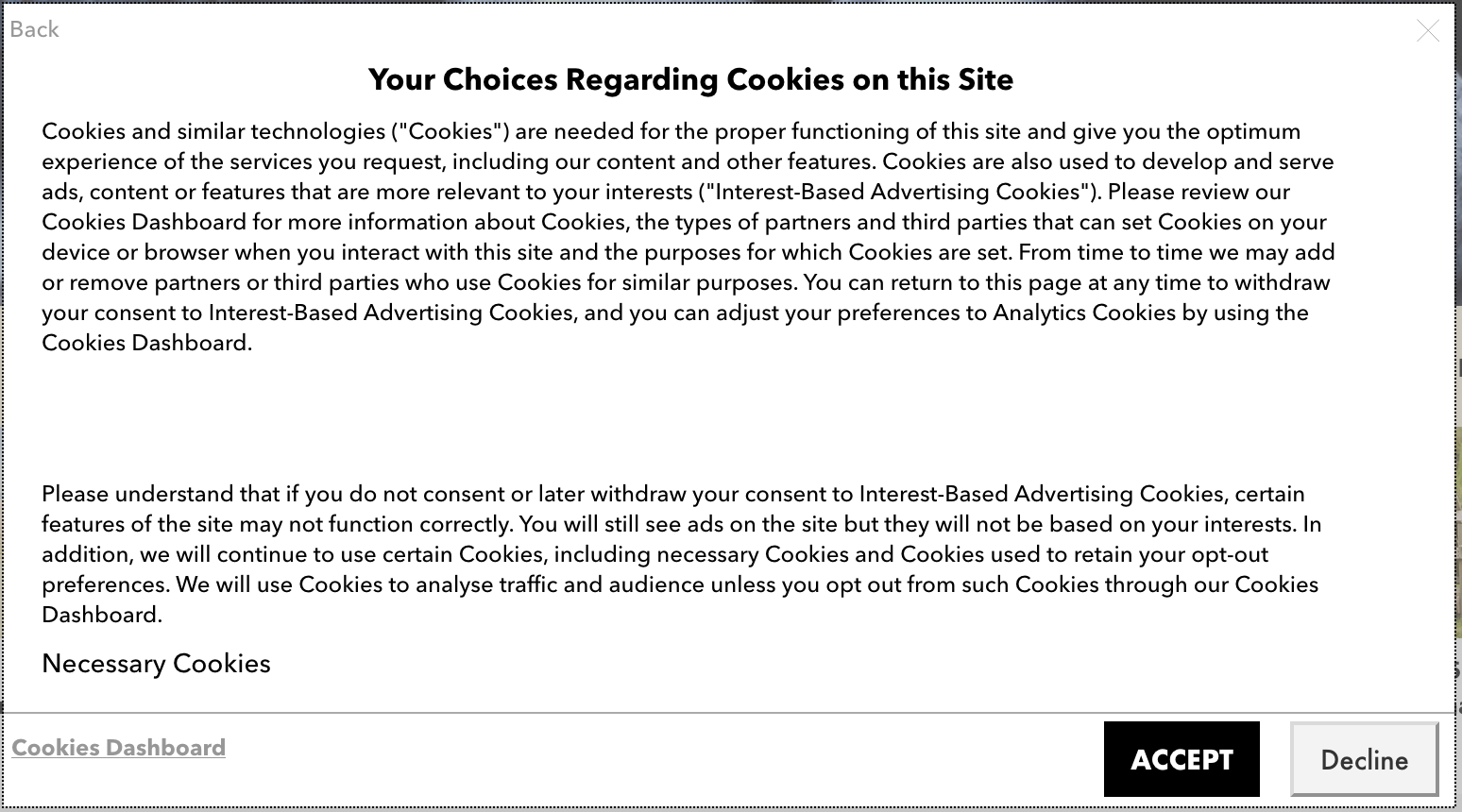}
\caption{\footnotesize An example of interface interference.  } \label{fig:interference}

\end{figure}

\paragraph{Interface interference} We encountered all forms of interface inferences, hidden information, pre-selection and aesthetic manipulation, described in \citep{Gray:2018}. The most clear example of interference is having the option to deny consent hidden by design as in the example in Figure~\ref{fig:interference} from CountryLiving.com: the consent notice first displayed (left) is a block of text; following the ``learn more" link leads to the actual notice (right) that allows the user to deny consent. 
 
\paragraph{Forced action} Forced action patterns are observed in consents implemented as screen popups that block the users from accessing the website and require the users to click on it before they can continue. Geo blocking can also be seen as a form of forced action. An example is given in Figure~\ref{fig:block}.
\begin{figure}[h!]
\centering
\includegraphics[width=0.8\textwidth]{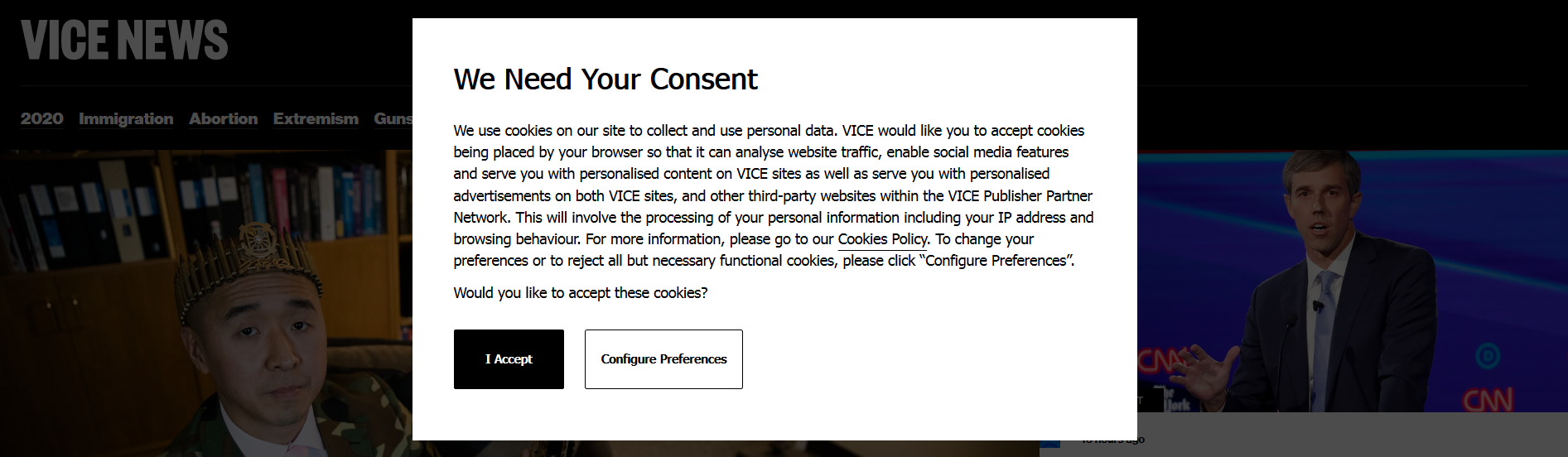}
\caption{\footnotesize An example of forced action from vice.com. } \label{fig:block}
\end{figure}

\subsection{Further Dark Pattern Observations}

After having conducted the initial data collection in July 2019 and having observed that at least one dark pattern was identified in all the visited websites, we conducted a  
second round of data collection and analysis  in April 2020,  revisiting all 300 web-pages. Using only the  (above mentioned) categories and descriptions of dark patterns from \citet{Gray:2018}, we attempted to identify which dark pattern occurs. The instruction provided to the reviewers specified to add a binary value (yes/no) for whether a dark pattern is present in the consent notice. A separate comment field was provided to allow for an argument why a dark pattern is considered as present.

\begin{figure}[h!]
	\centering
	\includegraphics[width=0.8\textwidth]{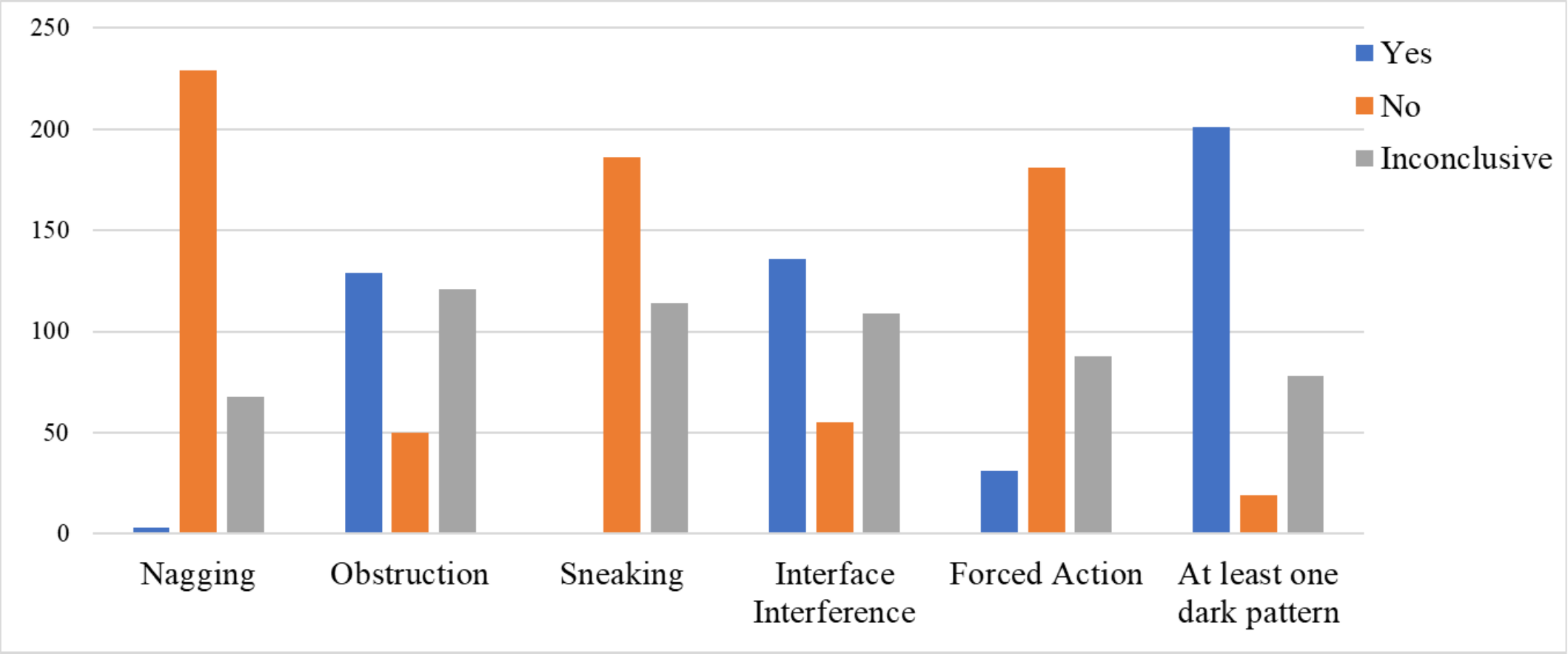}
	\caption{Results of dark patterns examination in websites} \label{fig:codingResultsBarGraph}
\end{figure}

Each site was examined by two researchers independently. The occurrences of dark patterns are summarized in Figure~\ref{fig:codingResultsBarGraph}. `Yes' and `No' indicate that both reviewers agreed on existence, or absence respectively,  of dark pattern. `Inconclusive' is a result of different answers from the two reviewers.

The results from the categorization show that the most patterns were Obstruction in 43\% (129 out of 300), and Interface Interference  occurring in 45,3 \% (136 out of 300), of the notices that contain dark patterns. 
Figure~\ref{fig:codingResultsBarGraph} shows considerable discrepancy in the classification of the presence of a given dark pattern with an inter-rater reliability of 67\%. This might imply that the inter-rater reliability is low, but it also is unsurprising given that the description of the dark patterns is insufficient for their characterisation. Using the rater's provided comments we attempted to discover the source of disagreements. There are several factors that can be considered relevant. Firstly, the dark pattern categories are sometimes overlapping even in the description of \citet{Gray:2018} and it can be difficult to decide which category different €˜`problems'€™ belong to. This could be seen in examples where one reviewer categorised a problem as obstruction while the other categorised the same problem as interface interference, and both have good arguments for their choice. Further, we could see that different reviewers had different focuses when looking for dark patterns. Some were more observant for design elements, while others focused more on wording and usability. When focusing on specific aspects one might also miss other features. This does not necessarily mean that the reviewers disagree conceptually, but rather that they have noticed different problems in the same cookie consent notice. 
 
 Figure~\ref{fig:codingResultsBarGraph} also shows that there are now websites that do not have a dark pattern in their cookie consent notice (right-most orange bar), in contrast to all surveyed websites exhibiting one in July 2019. This discrepancy can further be explained by the fact that some websites have changed their consent design (perhaps in response to CCPA coming to power). 

\section{Complexity of consent notices }\label{sec:complex}
\begin{wrapfigure}{r}{0.5\textwidth}{
	\begin{center}
	\includegraphics[width=0.5\textwidth]{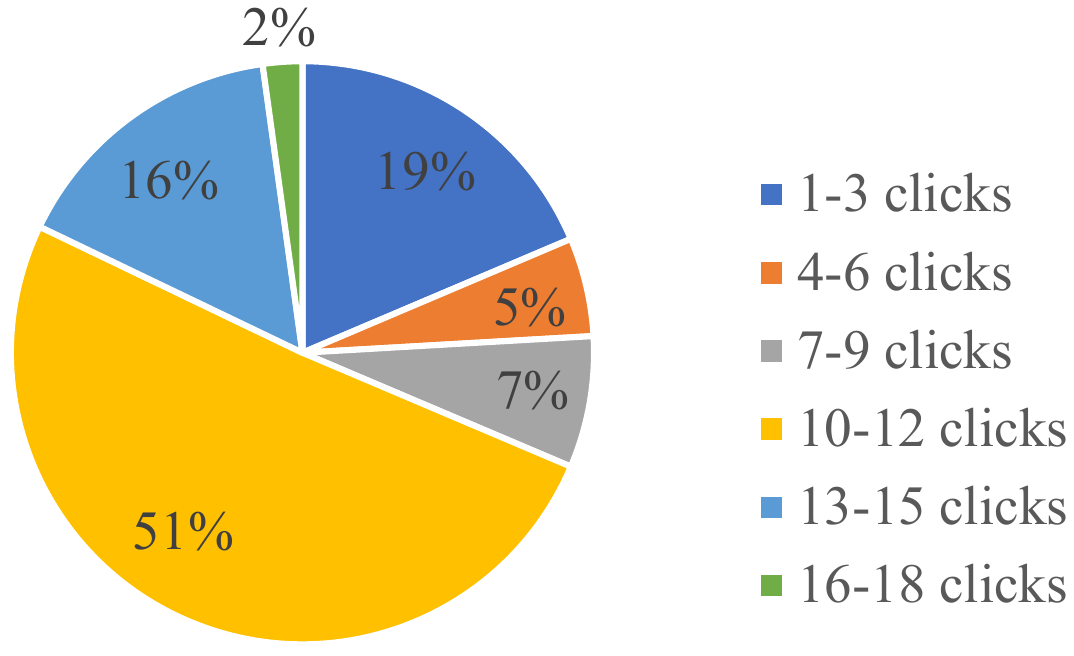}
	\caption{Number of clicks required to deny consent.} \label{fig:clicks-required-opt-out}
		\end{center}}
\end{wrapfigure}
The option to deny consent, directly or via browser settings, is available on 297 out of the 300 websites we surveyed in July 2019\footnote{As we observed, some of the websites have already updated their consent notice design}. However, these cookie consent notices vary in visibility and complexity of the denying consent option. The deny consent option is hidden in a scrollable or expandable area on 77 of the websites. In 220 of the websites the deny consent option does not have the same UI hierarchy as the accept option.

The number of clicks required to deny all types of cookies is collected as a proxy for the effort required to deny data collection consent.  We found that this number varies from 1, pressing one button on the displayed consent notice, up to 17, involving changing browser settings following the instruction provided on a separate website. Only 15 of the 300 websites provide a direct 1-click deny button, in contrast to all of them having 1-click accepts. As illustrated on Figure~\ref{fig:clicks-required-opt-out}, we can see that half of the websites require the user to make 10-12 clicks to opt out of all cookies.

\begin{table}[h!]
	\centering
	\resizebox{0.9\linewidth}{!}{%
	\begin{tabular}{|ll|ll|}
		\hline
		\multicolumn{2}{|l|}{Deny consent clickable description} & \multicolumn{2}{l|}{Cookie categories} \\
		\hline
		Word & Frequency & Word & Frequency \\ \hline
		Read more & 29 & Measurement  &  26\\
		Cookie Policy & 21 & Information storage and access  &  25 \\ 
		More Information & 17 & Personalization  &  24 \\ 
		Privacy Policy & 17  & Strictly necessary cookies  & 19 \\
		Learn More & 16 & Ad selection, delivery, reporting  &  19 \\ 
		Here  & 12  & Necessary  & 19 \\ 
		Cookie Settings & 6 & Unclassified  & 19\\ 
		Find Out More  & 6 & Content selection, delivery, reporting & 18 \\ 
		More info & 6 & Marketing & 18 \\ 
		Show purposes & 5 & Performance Cookies & 15 \\
		\hline

	\end{tabular}
	}
	\caption{Common words for deny consent clickable area and cookie categories}
	\label{table:deny-consent}
\end{table}

After translating all words into English, we have found that there are 121 different words being used in the clickable area of the deny consent. Most common words used are listed in the left hand-side column of  Table~\ref{table:deny-consent}. Words that frame the privacy friendly configuration directly such as ``deny'', ``reject", ``opt out", ``decline", are not widely used.

 Websites that instruct the user to  disable cookies on their browsers instead of offering this functionality within the consent widget, can be seen as not GDPR compliant\footnote{However, we acknowledge that there are technical challenges involved in adding configurable privacy settings on an existing website.}. 
The GDPR further stipulates that ``when the processing has multiple purposes, consent should be given for all of them". We observe that only 125 out of the 300 websites in the study listed the purpose of the cookies used.

Given that the study only collected data from news outlets and magazines, the purposes and categories of the data collection was expected to be similar among the surveyed web-pages. However, we found that the wording used to describe the data collection purposes is very diverse. We translated all stated purposes into English and found 201 distinct wordings for purposes of data collection in total. Each website on average has 4.7 purposes of cookies listed. We find it to be unreasonable to expect the user to familiarize themselves with all the different ways a type of cookies is called in different web-pages. We give the most commonly used cookie names in the right-hand side column of Table~\ref{table:deny-consent}.  

\begin{wraptable}{r}{0.45\textwidth}{
	\centering
	\resizebox{\linewidth}{!}{%
	\begin{tabular}{|l|l|}
	\hline
		Location of consent widget & Frequency \\ \hline
		Bottom, entire, less than 1/3 of page height & 131 \\
		Bottom, entire, around 1/3 page height & 23 \\
		Bottom, entire, around 1/2 of page height & 11 \\
		Bottom, left corner & 5 \\	
		Bottom, middle & 8\\
		Bottom, right corner & 7\\
		Middle of page, middle & 42 \\
		Others & 26\\
		Top, entire, less than 1/3 of page height & 30\\
		Top, entire 1/3 of page height & 4\\
		Top, middle & 3 \\
		Whole page &7 \\
		\hline
	\end{tabular}
	}
	\caption{Location of the consent widget.}
	\label{table:location-pop-up}
}
\end{wraptable}

Information about the screen position of the consent notice is summarized in Table~\ref{table:location-pop-up}. Exactly 60\% of the website display consent notices at the bottom part of the screen.   \citet{UtzDFSH2019} reported higher interaction rates for the notices displayed at the bottom and left side of the screen for desktop screens. They set up a web-page that offered the consent notice in various locations and compared the choices of 80.000 visitors.   On the other hand, \citet{MatteBS2019} ran a users study with 40 participants and did not observe any meaningful impact of the consent notice location on whether the user consented to cookies. Clearly more research is needed to understand the possible impact of the consent placement on the choice.

The content scrolling function was disabled when the cookie notice widget is active on 37 of the 300 websites. On 11 websites we noted the usage of a nagging dark pattern via the ``Are you sure?" type of messages all consent was denied.

Though the usage of third party  data privacy handling could improve the user experience by making the cookie preference configuration task done once and used everywhere, it is still very sparsely used -- we encounter it in only 5 out of 300 websites visited. After all the cookie consent is denied, 9 out of the 300 websites in our study stopped providing their services.

  \section{Discussion - refining dark pattern types in cookie consents}\label{sec:refine}
  
Using the insight from the related work and our own analysis we put forward the argument that more work is needed both on the end of regulators and UI specialists in protecting users from hindrance to exercise their rights by design. We look specifically in the case of consent elicitation. Before focusing on this argument in the rest of this section, we would like to point out that our  work illuminates a gap standards for what is now the cookie aspect of the data market industry. 

It is clear from our analysis, right hand-side column of  Table~\ref{table:deny-consent}, that the industry does not have a common language to address what are essentially the same type of cookies and cookie uses. The avoidance to use a ``negative" word for denial of consent  left hand-side column of  Table~\ref{table:deny-consent},  also introduces additional cognitive burden on the user.  There are over 1.5 billion websites\footnote{\url{https://www.internetlivestats.com/total-number-of-websites/}} at present. Even if there were only a thousand using different cookie descriptions, it would still  be difficult for an average user to read all of the cookie and purposes descriptions to understand what they are required to consent to. The data industry needs a standard of terminology that users can read and understand and rely on in consent elicitation.

It is clear that regulators need to include UI design requirements in regulations. Following the work of \citet{NouwensLVKK20}, the Danish Data Protection Authority has published a refinement to the guidelines on how consent notices should be designed. Specifically, among else, they include \citep{DataDK} the following pointers (our translation):

 \begin{itemize}
 \item The visitor to your website should be given an active option to allow their information to be processed
\item It should be clear for which different purposes you would like to process the information collected
\item It should be easy for the visitor to give consent for some purposes and not give consent for others
\item It should be easy not to give consent - even visually
\item In addition, you must be able to document what a visitor has consented to - and how the consent has been obtained.
\end{itemize} 
However, even these refined Danish guidelines such as these allow for a lot of room for interpretation and abuse. How should regulators state their requirements?    

One concrete requirement would be that the consent acceptance and denial use the same widget on the same level -  the option for acceptance should be next to the option for rejection in the same design.  Ideally regulators should explicitly exclude the use of dark patterns from the allowed design of consent elicitation widgets. However, these are not at all a trivial requirement. For the regulatory agencies to have power over how consent elicitation is implemented, they necessarily would have to be able to verify if the regulation is being followed. Given that there are over 1.5 billion web-pages, this verification has to be automatisable. Dark pattern detection is not. 

We were able to have a more in-depth insight into the ``ease'' of withholding data collection consent than perhaps \citep{UtzDFSH2019,NouwensLVKK20} because we analysed the consent notices ``manually''. Dark patterns are defined around the perceived {\em intention} to deceive the user. Intention, however is very hard to detect by automatic means. Furthermore, as it is demonstrated by our analysis given in Figure~\ref{fig:codingResultsBarGraph}, even human dark pattern detectors disagree on which dark pattern it is that is implemented. This is unsurprising, as \citet{Gray:2018} self argue  the ambiguity of his definitions and that they are overlapping. Ideally we need to further define the concept and types of dark patterns, perhaps for a specific context at the time, in such a way that: 
\begin{itemize}
    \item the features that characterise a dark pattern are clearly identifiable,
    \item the characterising features are easily computer-detectable.
\end{itemize}
Having such a dark pattern definition would enable regulators to automatically flag violators, which in turn, we expect,  would increase compliance. To contribute towards this goal, we highlight the patterns of misdirection we have identified, which can be seen as a refinement of the interface interference, obstruction and forced action patterns of \citet{Gray:2018}. Some of these have already been implicitly identified in the literature. 

\paragraph{Does not count} \citet{MatteBS2019} have indicated that although consent has not been given (yet or has been denied) data is collected anyway. This can be seen as a dark pattern and it is specific since it can only be computationally detected (by following what the browser does on the back-end.)

\paragraph{No choice.} All the included links and buttons lead to a page that either instructs to further pages that detail adjustments of browser settings, direct the user to contact third-party services (e.g. see ``Opting out" paragraph in \url{https://www.horseandrideruk.com/privacy-and-cookies/}),  or just ``explains'' cookies and purposes.

\paragraph{Multiple choice panels.} The user should be asked for consent in only one notice panel. This may appear obvious, but we have detected examples where the consent can be given in two panel, but the denial option (if offered at all) is given in only one, smaller, panel. Consider for example the Figure  of \url{https://www.manilatimes.net} accessed on May 3, 2020 where there is a center and bottom page consent panel.  

\begin{figure}[h!]
    \centering
    \includegraphics[width=0.9\textwidth]{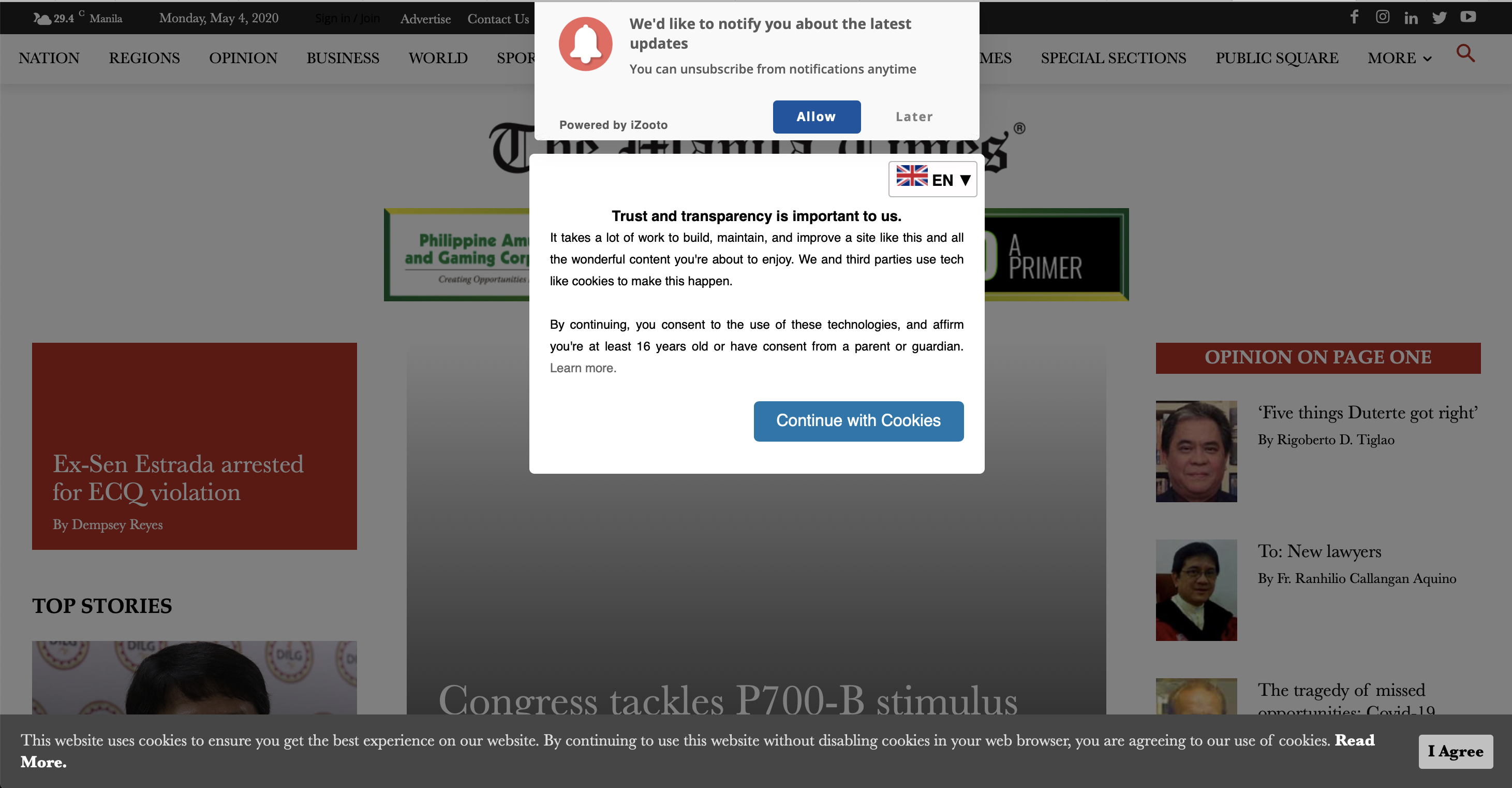}
    \caption{An example of multiple choice panels.}
    \label{fig:my_label}
\end{figure}
\begin{figure}[h!]
    \centering
    \includegraphics[width=0.3\textwidth]{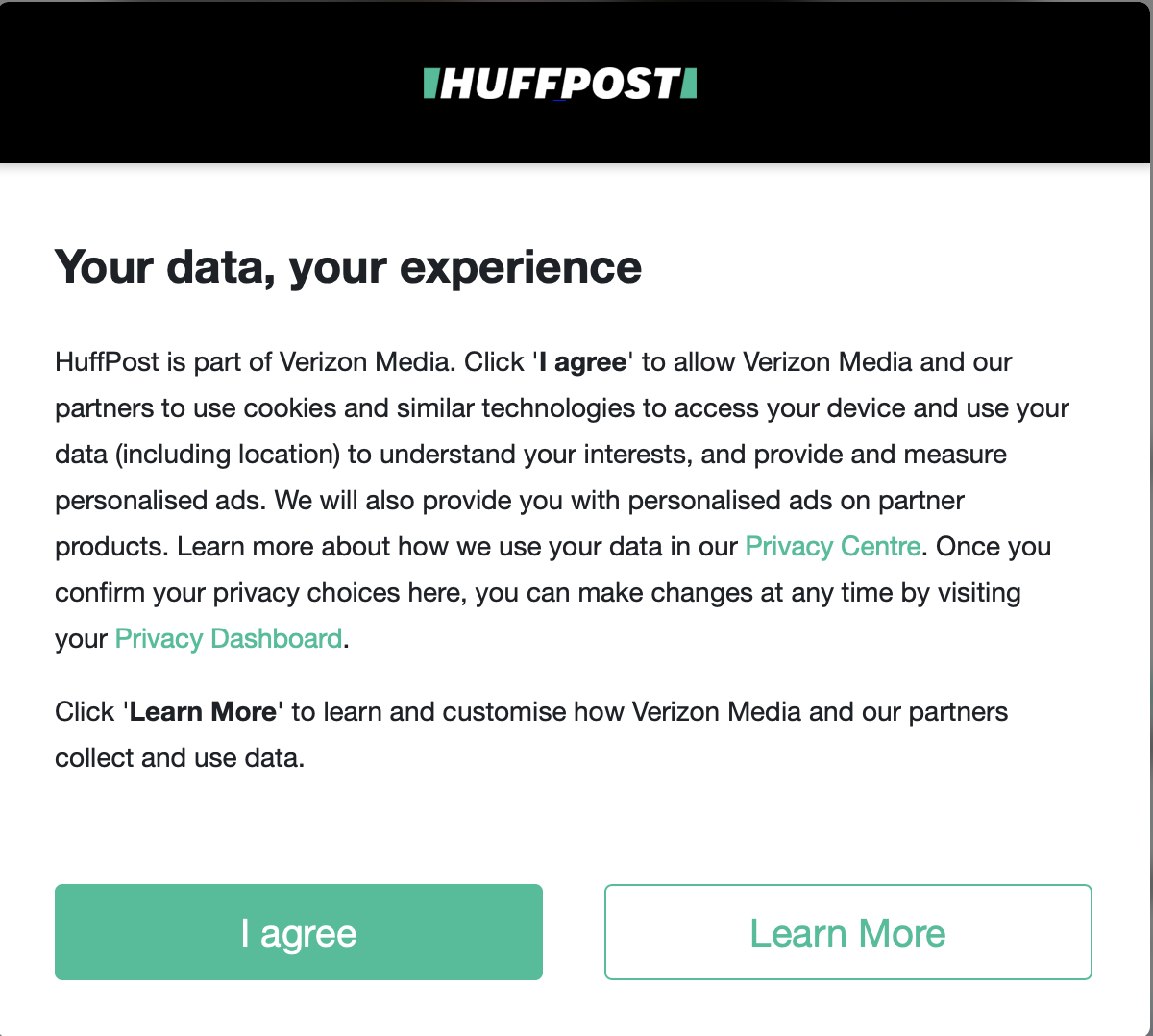}
     \includegraphics[width=0.3\textwidth]{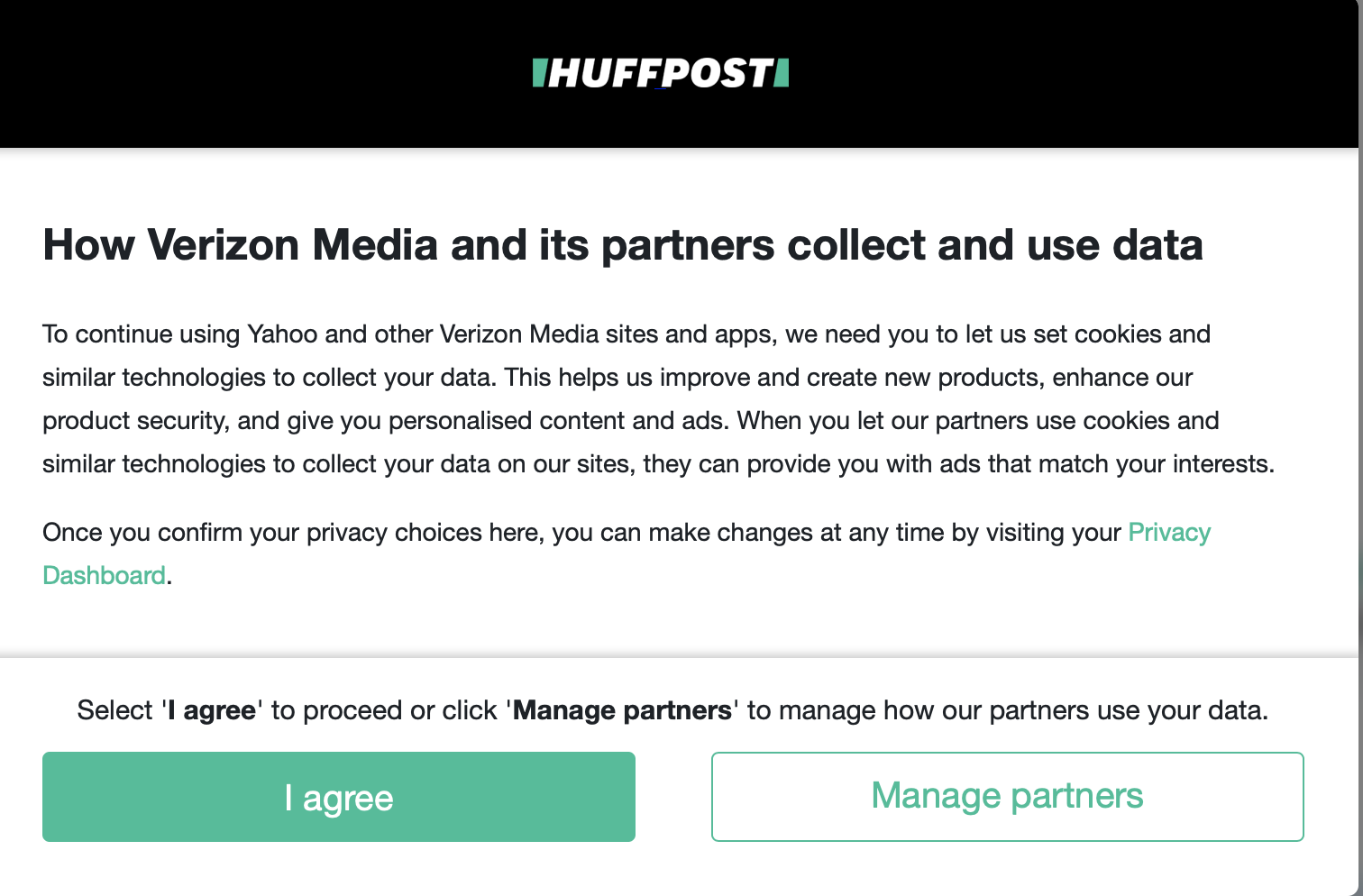}
      \includegraphics[width=0.3\textwidth]{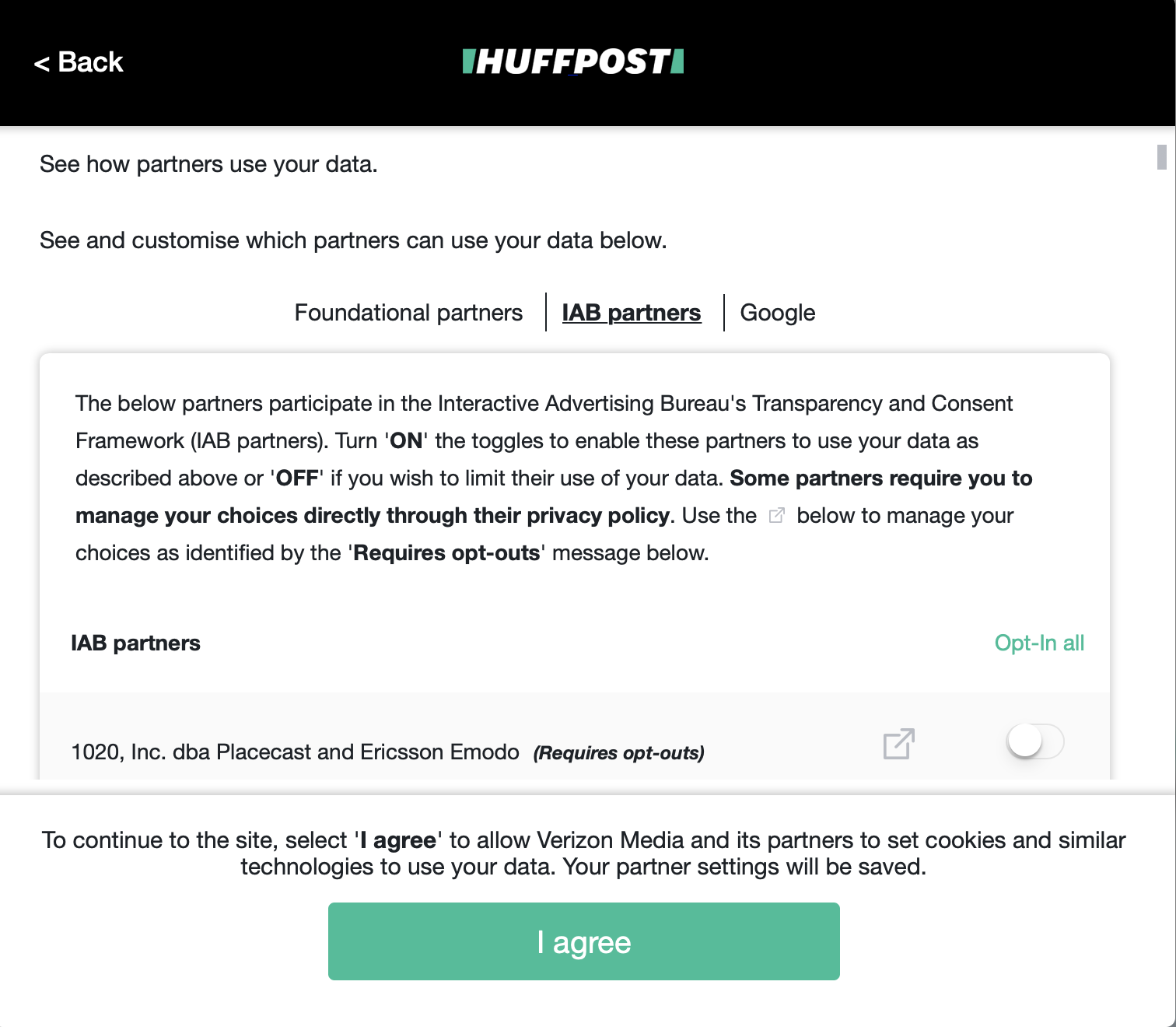}
    \caption{An example of a choice cascade \url{https://huffpost.com/} loading \url{consent.yahoo.com}}
    \label{fig:huffington}
\end{figure}

\paragraph{Choice cascade.} The denial of consent is only reached by following a number of links or buttons that offer more information. One example is given on Figure~\ref{fig:interference}: after following the learn more link the panel with the ``Decline" option appears. Another example is given in Figure~\ref{fig:huffington}: clicking ``learn more'' in the left most panel leads to the middle panel on which the user needs to click ``manage partners'' to reach the right panel where the middle tab reveals the consent opt-out sliders (if one scrolls down there is still no reject button and it is not clear whether the `I agree button' on this panel serves that role.)

\begin{figure}[h!]
    \centering
    \includegraphics[width=0.9\textwidth]{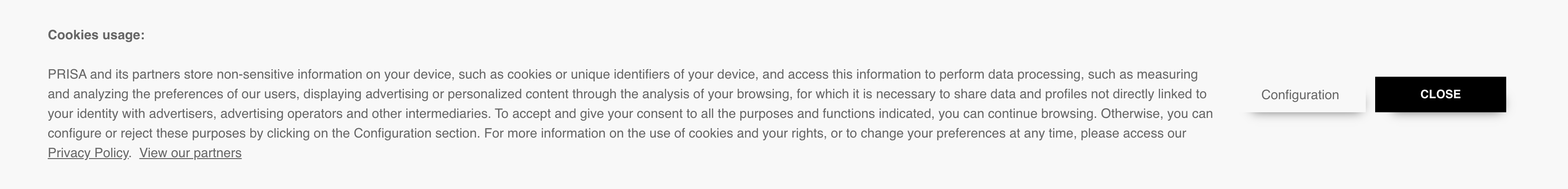}
\includegraphics[width=0.9\textwidth]{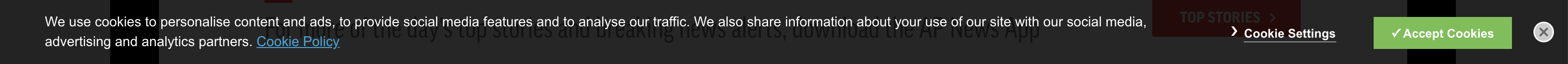}
     \includegraphics[width=0.9\textwidth]{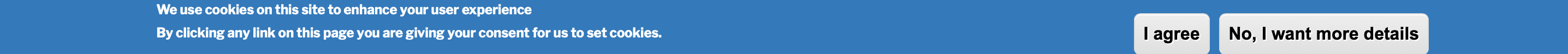}
    \caption{Examples of widget inequality. Top \url{https://english.elpais.com/}. Middle \url{ap.org}. Bottom \url{https://www.historytoday.com}.}
    \label{fig:inequal}
\end{figure}

\paragraph{Widget inequality} We have observed that while the execution of giving consent is made easy (bright button clearly labeled with a positive word), the denial of consent option, even if given in the same panel as the acceptance is given a different design. Examples range from a different web object altogether (link instead of a button),  single button for consent more than one for learning more and implicitly denial of consent, two buttons of which the one for consent is more noticeable (Figure~\ref{fig:inequal} top and middle),  to two identical button but with different functionality (Figure~\ref{fig:inequal} bottom).  

Either both choices lead to an explanation of cookies and purposes or neither should.  

\begin{figure}[h!]
    \centering
     \includegraphics[width=0.9\textwidth]{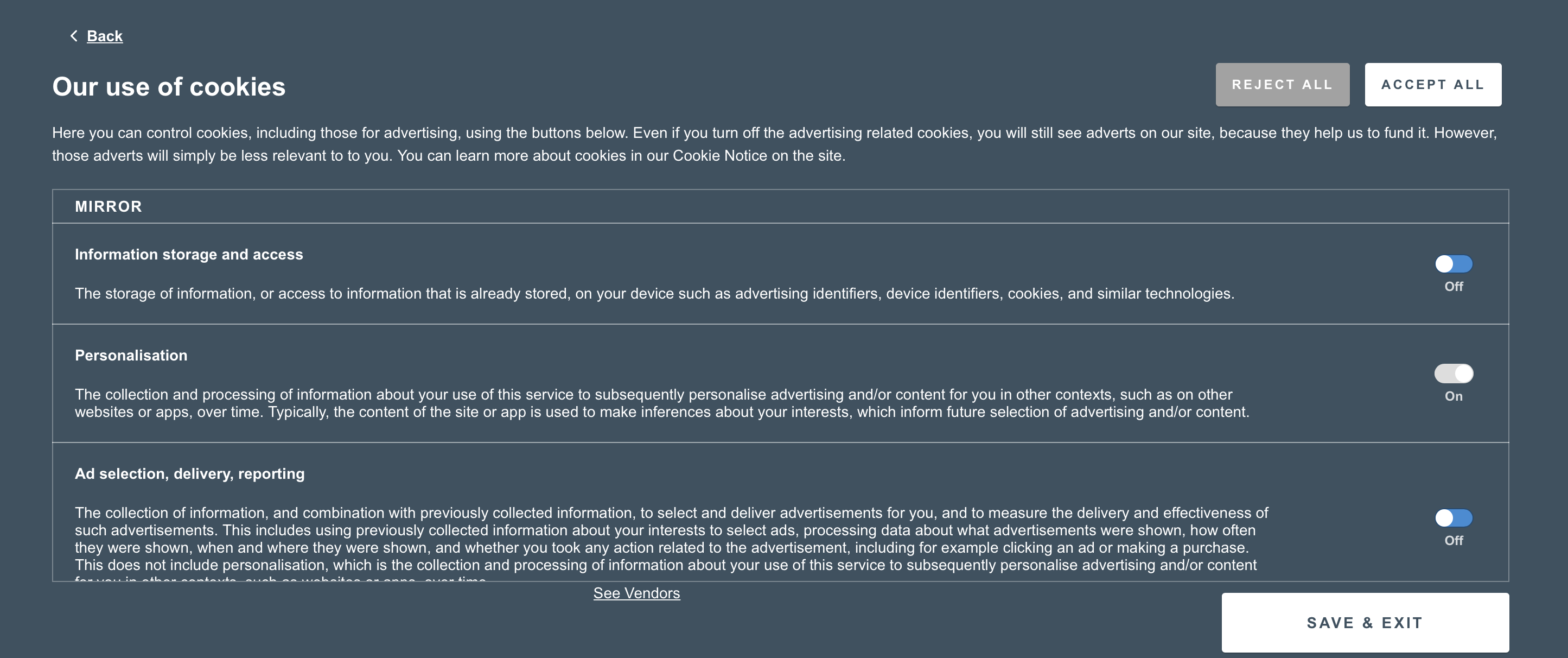}
    \caption{Examples of well labeled sliders   \url{https://www.mirror.co.uk}.}
    \label{fig:mirror}
\end{figure}

\paragraph{Unlabeled sliders} The consent notices uses sliders to allow users to consent or not to individual services, but it is not labeled which side of the panel is accept/on/active or which is reject/off/inactive. See for example the right most panel of Figure~{fig:huffington}.  For an example of well labeled sliders see Figure~\ref{fig:mirror}. 

\begin{figure}[h!]
    \centering
     \includegraphics[width=0.9\textwidth]{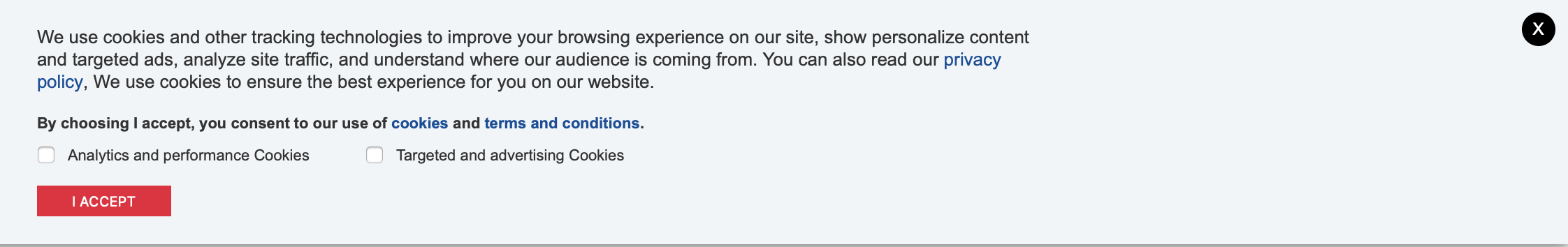}
    \caption{An example of an unmarked X from \url{https://economictimes.indiatimes.com}.}
    \label{fig:india}
\end{figure}

\paragraph{Unmarked X} When the panel has (usually) a top right ``x" widget, but the panel text does not explain whether clicking this ``x" counts as consent or denial of consent. See for example Figure~\ref{fig:india}.

\paragraph{No antonyms} Lastly, the use of clear words  such as ``I agree", ``I consent", ``yes" to label  consent option and not using their antonyms to label the denial of consent option, see for example Figure~\ref{fig:huffington},  in its own is a dark pattern.

 Lastly we put forward a  design suggestion for a cookie consent notice that we believe fulfills the requirements of the GDPR and does not have any dark patterns (old or new). This example builds on the design of \citet{CookieBot}, but with some adjustments.
 
 As the GDPR requires the user must give an informed consent for the different types of data collected. This requirement is fulfilled by giving the user the opportunity to select which data categories they want to consent to, if any. Only `Necessary' is pre-selected, which the GDPR allows as this is cookies that are needed for the website to function. Further, all cookie options are displayed at the first page: The user can both accept or deny all cookies with one click. If the user wants to allow only specific cookie categories this is done with a few clicks. All the button have the same size and color as not to indicate to the user that one alternative is more correct than the others. The wordings on the different buttons are made as clear as possible not to confuse the user about their purpose. 
 If the user wants to read more about the cookies or the different cookie categories they can do so by clicking `More information'.
 
 It is important that the cookie consent notice don't force the user to make a choice. The cookie consent notice must therefore be placed accordingly. Lastly, it is important that the user know how to change the settings later on. The cookie consent notice should include this type of information and the website should make this information accessible in their website. 
 
\begin{figure}[h!]
\centering
\includegraphics[width=0.8\textwidth]{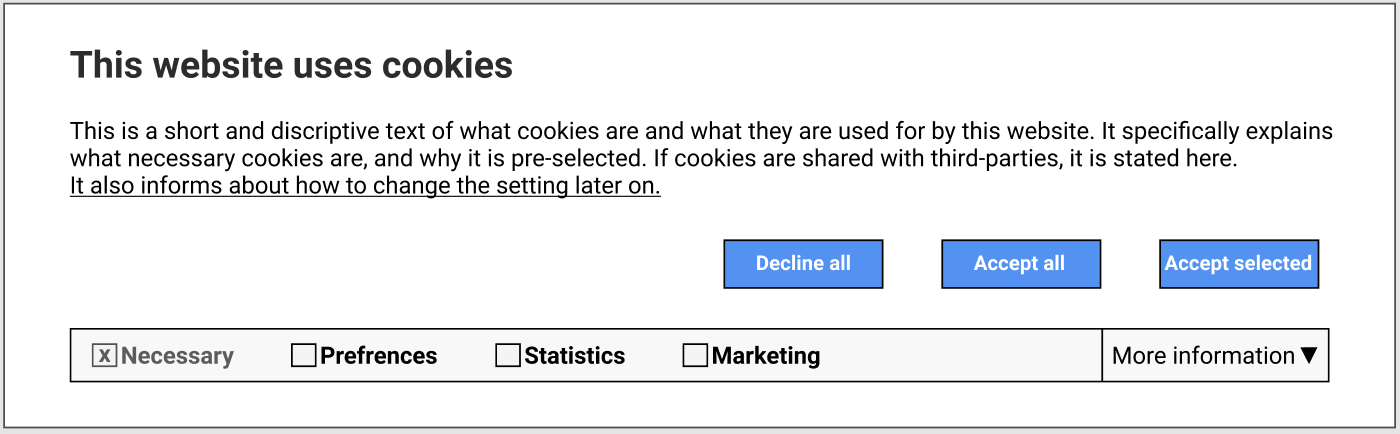}
\caption{\footnotesize An example of cookie consent notice. } \label{fig:block}
\end{figure}

\section{Summary}\label{sec:summary}

In this paper we have presented an analysis of 300 consent notices from a selection of online news outlets. The analysis shows that most of them (297) use dark patterns when eliciting consent from their users. Further we categorized the prevalence of the different dark patterns as defined by Gray \citep{Gray:2018} . Based on the analysis, we identified and described eight specific patterns used in the analyzed cookie consent notices. These patterns can be used when evaluating consent notices to make sure they are not a circumvention by design of the intentions of privacy regulations.

Unlike  \citep{UtzDFSH2019,MatteBS2019,NouwensLVKK20}  who scraped the data on the cookie notices that were analyzed,  we collected our data manually. This resulted in a smaller, and richer data set, and allowed us to study features of user interaction in design that cannot be easily automatically detected. 

In contrast to \citet{UtzDFSH2019} and \citet{NouwensLVKK20}  we did not run a user study to see what effect the usage of dark patterns have and focused on a detailed analysis of the implemented designs for  the consent notice. We however look for more features of design such as type of dark pattern and complexity of interaction, expressed in amount of effort required to opt out and richness of textual descriptions for instructions and purposes.

We discuss some of the  limitations of ours, and the work \citep{UtzDFSH2019,MatteBS2019,NouwensLVKK20} done so far. 
 We conducted our survey on browsers running on laptops. 
 The size of the consent notice is not always possible to scale with the size of the screen, as that would make the text of the notice illegible. As a result, the consent notice that would cover a third of a screen on a laptop browser, would cover the entire screen on a mobile device which might change the likelihood of consent \citep{UtzDFSH2019}.  
We analyzed  each consent notice from the perspective of a user that sees it for the first time. Repeated visits could reveal further dark patterns, in particular nudging.   In future work we intend to explore such repeated behavior of web services. Web-pages for which data collection consent was denied can persist on requesting it with each new open sub-page or at each subsequent visit despite ``functionality" or ``necessary" cookies being used that could store the denial of consent information. It is also possible that the design of the interface changes for subsequent visitors, or it is personalized. We also did not compare whether the ``processing time'' for the choice and accessing the web-page (when access is blocked prior to choice) is the same for the accept and reject options. We have noticed that sometimes after opting out from consent, processing takes noticeably long. There are reasons to believe that there might be a whole new type of dark patterns that reveal themselves after the consent panel has been interacted with.   

We did not investigate whether cookie consents are geo-dependent, specifically if citizens accessing the same web-page from a non European country are still offered the opportunity to opt out from data collection and if the consent design differs. 

Analyzing our work and related work,  we made an attempted to refine the dark pattern concept by identifying dark patterns that we have encountered in cookie consent notices. Our list is not intended to be exhaustive, merely a first step towards helping regulators and experts identify that undesirable design choices have been made. We stipulated that a useful dark pattern definition is such whose characterising features are clearly and automatically identifiable. In our future work we intend to assert that our refined dark pattern types satisfy this criteria of usefulness by building a crawler that identifies these dark patterns in cookie consent notices. Our data set allows for each of the 300 considered consents to be labeled with the new dark patterns allowing us to explore using machine learning for dark pattern identification. 
 \bibliographystyle{ACM-Reference-Format}
\bibliography{moralUX}


\begin{thebibliography}{19}


\ifx \showCODEN    \undefined \def \showCODEN     #1{\unskip}     \fi
\ifx \showDOI      \undefined \def \showDOI       #1{#1}\fi
\ifx \showISBNx    \undefined \def \showISBNx     #1{\unskip}     \fi
\ifx \showISBNxiii \undefined \def \showISBNxiii  #1{\unskip}     \fi
\ifx \showISSN     \undefined \def \showISSN      #1{\unskip}     \fi
\ifx \showLCCN     \undefined \def \showLCCN      #1{\unskip}     \fi
\ifx \shownote     \undefined \def \shownote      #1{#1}          \fi
\ifx \showarticletitle \undefined \def \showarticletitle #1{#1}   \fi
\ifx \showURL      \undefined \def \showURL       {\relax}        \fi
\providecommand\bibfield[2]{#2}
\providecommand\bibinfo[2]{#2}
\providecommand\natexlab[1]{#1}
\providecommand\showeprint[2][]{arXiv:#2}

\bibitem[\protect\citeauthoryear{Borders}{Borders}{2020}]%
        {Rank2020}
\bibfield{author}{\bibinfo{person}{Reporters~Without Borders}.}
  \bibinfo{year}{2020}\natexlab{}.
\newblock \bibinfo{booktitle}{\emph{World Press Freedom Index}}.
\newblock Supported by the Adessium Foundation.
\newblock
\urldef\tempurl%
\url{https://rsf.org/en/ranking}
\showURL{%
\tempurl}


\bibitem[\protect\citeauthoryear{Brignull and Darlington}{Brignull and
  Darlington}{2010}]%
        {Brignull2010}
\bibfield{author}{\bibinfo{person}{Harry Brignull} {and}
  \bibinfo{person}{Alexander Darlington}.} \bibinfo{year}{2010}\natexlab{}.
\newblock \bibinfo{booktitle}{\emph{Getting round GDPR with dark patterns. A
  case study: Techradar}}.
\newblock
\urldef\tempurl%
\url{https://www.darkpatterns.org/}
\showURL{%
\tempurl}


\bibitem[\protect\citeauthoryear{CookieBot}{CookieBot}{2020}]%
        {CookieBot}
\bibfield{author}{\bibinfo{person}{CookieBot}.}
  \bibinfo{year}{2020}\natexlab{}.
\newblock \bibinfo{booktitle}{\emph{Solution for Compliant Use of Cookies and
  Online Tracking}}.
\newblock Cybot Denmark.
\newblock
\urldef\tempurl%
\url{https://www.cookiebot.com/}
\showURL{%
\tempurl}


\bibitem[\protect\citeauthoryear{Denmark}{Denmark}{2020}]%
        {DataDK}
\bibfield{author}{\bibinfo{person}{Datatylsinet Denmark}.}
  \bibinfo{year}{2020}\natexlab{}.
\newblock \bibinfo{booktitle}{\emph{Nye retningslinjer om behandling af
  personoplysninger om hjemmesidebes{\o}gende}}.
\newblock
\urldef\tempurl%
\url{https://www.datatilsynet.dk/presse-og-nyheder/nyhedsarkiv/2020/feb/nye-retningslinjer-om-behandling-af-personoplysninger-om-hjemmesidebesoegende/}
\showURL{%
\tempurl}


\bibitem[\protect\citeauthoryear{Ducato and Marique}{Ducato and
  Marique}{2019}]%
        {DucatoE2019}
\bibfield{author}{\bibinfo{person}{Rossana Ducato} {and}
  \bibinfo{person}{Enguerrand Marique}.} \bibinfo{year}{2019}\natexlab{}.
\newblock \bibinfo{title}{Come to the Dark Side: We Have Patterns. Choice
  Architecture and Design for (Un)Informed Consent}.
\newblock
\newblock
\urldef\tempurl%
\url{https://ssrn.com/abstract=3365952}
\showURL{%
\tempurl}


\bibitem[\protect\citeauthoryear{Forbrukerr{\aa}det}{Forbrukerr{\aa}det}{2018}]%
        {ForbrukerradeNO}
\bibfield{author}{\bibinfo{person}{Norway Forbrukerr{\aa}det}.}
  \bibinfo{year}{2018}\natexlab{}.
\newblock \bibinfo{booktitle}{\emph{DECEIVED BY DESIGN: How tech companies use
  dark patterns to discourage us from exercising our rights to privacy}}.
\newblock Forbrukerr{\aa}det, Norway.
\newblock
\urldef\tempurl%
\url{https://fil.forbrukerradet.no/wp-content/uploads/2018/06/2018-06-27-deceived-by-design-final.pdf}
\showURL{%
\tempurl}


\bibitem[\protect\citeauthoryear{Garfinkel and Cox}{Garfinkel and Cox}{2009}]%
        {Garfinkel2009}
\bibfield{author}{\bibinfo{person}{Simson~L. Garfinkel} {and}
  \bibinfo{person}{David~Alexander Cox}.} \bibinfo{year}{2009}\natexlab{}.
\newblock \showarticletitle{Finding and Archiving the Internet Footprint}. In
  \bibinfo{booktitle}{\emph{First Digital Lives Research Conference: Personal
  Digital Archives for the 21st Century, London, England, 9?11 February 2009}}.
\newblock
\newblock
\shownote{\url{https://simson.net/clips/academic/2009.BL.InternetFootprint.pdf}.}


\bibitem[\protect\citeauthoryear{Gray, Kou, Battles, Hoggatt, and Toombs}{Gray
  et~al\mbox{.}}{2018}]%
        {Gray:2018}
\bibfield{author}{\bibinfo{person}{Colin~M. Gray}, \bibinfo{person}{Yubo Kou},
  \bibinfo{person}{Bryan Battles}, \bibinfo{person}{Joseph Hoggatt}, {and}
  \bibinfo{person}{Austin~L. Toombs}.} \bibinfo{year}{2018}\natexlab{}.
\newblock \showarticletitle{The Dark (Patterns) Side of UX Design}. In
  \bibinfo{booktitle}{\emph{Proceedings of the 2018 CHI Conference on Human
  Factors in Computing Systems}} \emph{(\bibinfo{series}{CHI '18})}.
  \bibinfo{publisher}{ACM}, \bibinfo{address}{New York, NY, USA}, Article
  \bibinfo{articleno}{534}, \bibinfo{numpages}{14}~pages.
\newblock
\showISBNx{978-1-4503-5620-6}
\urldef\tempurl%
\url{https://doi.org/10.1145/3173574.3174108}
\showDOI{\tempurl}


\bibitem[\protect\citeauthoryear{Infromation}{Infromation}{2018}]%
        {CCPA}
\bibfield{author}{\bibinfo{person}{California~Legislative Infromation}.}
  \bibinfo{year}{2018}\natexlab{}.
\newblock \bibinfo{booktitle}{\emph{Assembly Bill No. 375, CHAPTER 55,
  Legislative Councel's Digest}}.
\newblock The state of California.
\newblock
\urldef\tempurl%
\url{https://leginfo.legislature.ca.gov/faces/billTextClient.xhtml?bill_id=201720180AB375}
\showURL{%
\tempurl}


\bibitem[\protect\citeauthoryear{Leit\~{a}o and Jakobsen}{Leit\~{a}o and
  Jakobsen}{2018}]%
        {Leitao:2018}
\bibfield{author}{\bibinfo{person}{Roxanne Leit\~{a}o} {and}
  \bibinfo{person}{Filip Jakobsen}.} \bibinfo{year}{2018}\natexlab{}.
\newblock \showarticletitle{A Survey on User-Interface Design Strategies to
  Address Online Bias}. In \bibinfo{booktitle}{\emph{Extended Abstracts of the
  2018 CHI Conference on Human Factors in Computing Systems}}
  \emph{(\bibinfo{series}{CHI EA '18})}. \bibinfo{publisher}{ACM},
  \bibinfo{address}{New York, NY, USA}, Article \bibinfo{articleno}{LBW084},
  \bibinfo{numpages}{6}~pages.
\newblock
\showISBNx{978-1-4503-5621-3}
\urldef\tempurl%
\url{https://doi.org/10.1145/3170427.3188567}
\showDOI{\tempurl}


\bibitem[\protect\citeauthoryear{Matte, Bielova, and Santos}{Matte
  et~al\mbox{.}}{2019a}]%
        {CookieGlasses}
\bibfield{author}{\bibinfo{person}{C\'{e}lestin Matte},
  \bibinfo{person}{Nataliia Bielova}, {and} \bibinfo{person}{Cristiana
  Santos}.} \bibinfo{year}{2019}\natexlab{a}.
\newblock \bibinfo{title}{Cookie Glasses}.
\newblock
  \bibinfo{howpublished}{\url{https://github.com/Perdu/Cookie-Glasses}}.
\newblock


\bibitem[\protect\citeauthoryear{Matte, Bielova, and Santos}{Matte
  et~al\mbox{.}}{2019b}]%
        {MatteBS2019}
\bibfield{author}{\bibinfo{person}{C{\'{e}}lestin Matte},
  \bibinfo{person}{Nataliia Bielova}, {and} \bibinfo{person}{Cristiana
  Santos}.} \bibinfo{year}{2019}\natexlab{b}.
\newblock \showarticletitle{Do Cookie Banners Respect my Choice? Measuring
  Legal Compliance of Banners from {IAB} Europe's Transparency and Consent
  Framework}.
\newblock \bibinfo{journal}{\emph{CoRR}}  \bibinfo{volume}{abs/1911.09964}
  (\bibinfo{year}{2019}).
\newblock
\showeprint[arxiv]{1911.09964}
\urldef\tempurl%
\url{http://arxiv.org/abs/1911.09964}
\showURL{%
\tempurl}


\bibitem[\protect\citeauthoryear{Newman, Fletcher, Kalogeropoulos, and
  Nielsen}{Newman et~al\mbox{.}}{2019}]%
        {Reuters2019}
\bibfield{author}{\bibinfo{person}{Nic Newman}, \bibinfo{person}{Richard
  Fletcher}, \bibinfo{person}{Antonis Kalogeropoulos}, {and}
  \bibinfo{person}{Rasmus~Kleis Nielsen}.} \bibinfo{year}{2019}\natexlab{}.
\newblock \bibinfo{booktitle}{\emph{Reuters Institute Digital News Report
  2019}}.
\newblock Reuters Institute and the University of Oxford.
\newblock
\urldef\tempurl%
\url{https://reutersinstitute.politics.ox.ac.uk/sites/default/files/2019-06/DNR_2019_FINAL_1.pdf}
\showURL{%
\tempurl}


\bibitem[\protect\citeauthoryear{Nouwens, Liccardi, Veale, Karger, and
  Kagal}{Nouwens et~al\mbox{.}}{2020a}]%
        {ConMAT}
\bibfield{author}{\bibinfo{person}{Midas Nouwens}, \bibinfo{person}{Ilaria
  Liccardi}, \bibinfo{person}{Michael Veale}, \bibinfo{person}{David Karger},
  {and} \bibinfo{person}{Lalana Kagal}.} \bibinfo{year}{2020}\natexlab{a}.
\newblock \bibinfo{title}{Consent-O-Matic}.
\newblock
  \bibinfo{howpublished}{\url{https://github.com/cavi-au/Consent-O-Matic}}.
\newblock


\bibitem[\protect\citeauthoryear{Nouwens, Liccardi, Veale, Karger, and
  Kagal}{Nouwens et~al\mbox{.}}{2020b}]%
        {NouwensLVKK20}
\bibfield{author}{\bibinfo{person}{Midas Nouwens}, \bibinfo{person}{Ilaria
  Liccardi}, \bibinfo{person}{Michael Veale}, \bibinfo{person}{David Karger},
  {and} \bibinfo{person}{Lalana Kagal}.} \bibinfo{year}{2020}\natexlab{b}.
\newblock \showarticletitle{Dark Patterns after the {GDPR:} Scraping Consent
  Pop-ups and Demonstrating their Influence}.
\newblock \bibinfo{journal}{\emph{CoRR}}  \bibinfo{volume}{abs/2001.02479}
  (\bibinfo{year}{2020}).
\newblock
\showeprint[arxiv]{2001.02479}
\urldef\tempurl%
\url{http://arxiv.org/abs/2001.02479}
\showURL{%
\tempurl}


\bibitem[\protect\citeauthoryear{Parlament and Council}{Parlament and
  Council}{2016}]%
        {GDPR}
\bibfield{author}{\bibinfo{person}{European Parlament} {and}
  \bibinfo{person}{Council}.} \bibinfo{year}{2016}\natexlab{}.
\newblock \bibinfo{booktitle}{\emph{Regulation (EU) 2016/679 of the European
  Parliament and of the Council of 27 April 2016 on the protection of natural
  persons with regard to the processing of personal data and on the free
  movement of such data, and repealing Directive 95/46/EC (General Data
  Protection Regulation)}}.
\newblock EU.
\newblock
\urldef\tempurl%
\url{http://data.europa.eu/eli/reg/2016/679/oj}
\showURL{%
\tempurl}


\bibitem[\protect\citeauthoryear{Paternoster}{Paternoster}{2018}]%
        {Paternoster2018}
\bibfield{author}{\bibinfo{person}{Leon Paternoster}.}
  \bibinfo{year}{2018}\natexlab{}.
\newblock \bibinfo{booktitle}{\emph{Getting round GDPR with dark patterns. A
  case study: Techradar}}.
\newblock Suffolk Libraries.
\newblock
\urldef\tempurl%
\url{https://www.leonpaternoster.com/posts/techradar-gdpr/}
\showURL{%
\tempurl}


\bibitem[\protect\citeauthoryear{Pedersen and Dyrkolbotn}{Pedersen and
  Dyrkolbotn}{2018}]%
        {NIK2018}
\bibfield{author}{\bibinfo{person}{Truls Pedersen} {and}
  \bibinfo{person}{Sjur~Kristoffer Dyrkolbotn}.}
  \bibinfo{year}{2018}\natexlab{}.
\newblock \showarticletitle{The legally mandated approximate language about
  AI}.
\newblock \bibinfo{journal}{\emph{Norsk Informatikkonferanse}}
  (\bibinfo{year}{2018}).
\newblock
\showISSN{1892-0721}
\urldef\tempurl%
\url{https://ojs.bibsys.no/index.php/NIK/article/view/506}
\showURL{%
\tempurl}


\bibitem[\protect\citeauthoryear{Utz, Degeling, Fahl, Schaub, and Holz}{Utz
  et~al\mbox{.}}{2019}]%
        {UtzDFSH2019}
\bibfield{author}{\bibinfo{person}{Christine Utz}, \bibinfo{person}{Martin
  Degeling}, \bibinfo{person}{Sascha Fahl}, \bibinfo{person}{Florian Schaub},
  {and} \bibinfo{person}{Thorsten Holz}.} \bibinfo{year}{2019}\natexlab{}.
\newblock \showarticletitle{(Un)Informed Consent: Studying GDPR Consent Notices
  in the Field}. In \bibinfo{booktitle}{\emph{Proceedings of the 2019 ACM
  SIGSAC Conference on Computer and Communications Security}}
  \emph{(\bibinfo{series}{CCS ’19})}. \bibinfo{publisher}{Association for
  Computing Machinery}, \bibinfo{address}{New York, NY, USA},
  \bibinfo{pages}{973–990}.
\newblock
\showISBNx{9781450367479}
\urldef\tempurl%
\url{https://doi.org/10.1145/3319535.3354212}
\showDOI{\tempurl}


\end{thebibliography}

\end{document}